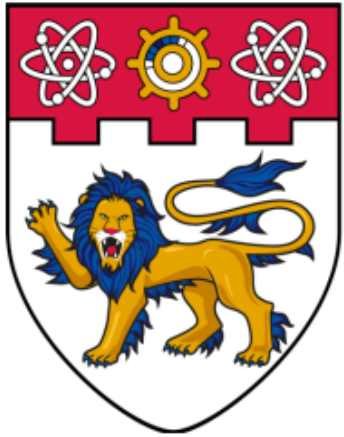

**CCDS24-0500**

**Low-power Wireless Network with Real-Time Guarantees for Edge-Cloud Applications**

**Don Tan Kiang Yong**

**U2120088K**

**Supervisor: A/P Arvind Easwaran**

**College of Computing and Data Science**

A final year project report presented to Nanyang Technological University in partial fulfilment of the Requirements for the Degree of Bachelor of Computing (Computer Science)

**2025**

# Abstract

The goal of this project is to explore the feasibility of building a scalable & easy-to-deploy real-time LoRa testbed, made from multiple units of Raspberry Pi (RPI), where each RPI manages its own set of LoRa radios. This project is motivated by the lack of concrete large-scale LoRa testbeds that effectively integrate LoRa communications into the real-time world. The paper introduces how the idea of using RPI came about and why it should work in theory. The paper then carries out experiments on a component of the large-scale testbed, to evaluate the feasibility of the said component based on performance metrics such as RSSI, SNR, PLR and the ability to carry out millisecond-accurate transmissions. The performance metrics are also used to explore the impact of using different combinations of spread factors and transmission frequencies, as well as making comparisons between time-division multiple access (TDMA) and carrier-sense multiple access (CSMA) approaches. The results show that with the right parameters configured, the system can achieve stable and low-latency communications, proving some feasibility to operate under real-time situations. Future work includes giving each RPI control over more radios, carrying out true parallel transmissions, and finally integrating multiple RPIs for a more complete large-scale real-time LoRa testbed.

# Acknowledgement

I would like to express my deepest gratitude to Amalinda Gamage for his proactive support and guidance throughout the first half of my final year project, before he moved onto the next chapter of his life at NUS. During the start, where everything was still uncertain for me, he imparted many valuable pieces of knowledge and resources, equipped me with tools to search for the answers by myself rather than giving it to me. Despite his busy schedule at another location, he still finds time to catch up with the progress of the project and give insights. Every interaction with him has immensely helped shape my academic understanding of this project and its outcome. I would also like to express my gratitude to Soumya, a Ph.D student, for his extensive guidance and help, especially towards the second half of my project. Even though he had lessons to teach and papers to write, he still makes sure that I am able to find him on campus for any questions or doubts. This project would not have been successful without the contributions of either.

# Table of Contents

# List of Figures

# 1. Introduction

## 1.1. Background

In recent years, Low Power Wide Area Networks (LPWAN), particularly LoRa (Long Range), have garnered much attention due to their long-range low-power communication capabilities. As a result of these two main features, LoRa becomes a highly attractive option for a variety of applications, such as real-time systems. When considering real-time systems, minimal delay is of utmost importance. Although communication technologies like Wi-Fi and 5G easily offer minimal delays, they are often limited by their range. Applications such as chemical processing plants, where gas leaks or undesirable pressure, temperature and pH levels can lead to catastrophic outcomes, and industrial automation, where a minor mispositioning of an object could lead to things breaking, could use a sensor to detect and prevent unwanted circumstances by sending fast and reliable updates over vast distances to ensure efficiency and minimal operational risks. These are scenarios where LoRa outshine other communication technologies in bringing value to real-time systems.

## 1.2. Challenges in Scaling Real-Time LoRa Testbeds

However, creating a large-scale real-time LoRa testbed introduces some challenges. Despite the growing interest in utilizing LoRa for real-time applications, there is still a lack of large-scale testbeds in literature [20]. There is a lack of data on how the testbed will perform when deployed at larger scales. Hence, there is uncertainty regarding their viability and the optimal strategies for implementing them. Additionally, as the size of the testbed increases, so does the complexity of managing the end devices, synchronizing the system, and the handling of backhaul data. A larger testbed entails more devices being involved, more extensive data collection, and more complex synchronization efforts. These factors cause an exponential increase in the time required to test and evaluate the system. This makes the testing process impractical because it would take too much time. Even more so for real-time applications because data transmissions must be rigorously tested to ensure reliability and performance to the milliseconds.

# 2. Motivation

## 2.1. Obstacles in Building a Scalable Real-Time LoRa Testbed

These challenges give rise to the main motivation of this paper: Can we build a scalable and easy-to-deploy real-time LoRa testbed? Before giving this question a proper response, there is a need to consider the obstacles ahead. A main bulk of challenges will come from the concept of large-scale alone.

### 2.1.1. Uploading Code

A large number of end devices results in impractical manual intervention when it comes to uploading codes and extracting backhaul data. In a simple LoRa system, there is a need to upload codes into the end devices because these codes will be what determines how the nodes will behave within the network. For example, for a LoRa radio, the code can determine how often the radio will be making transmissions. For a regular sensor, the code can instruct the sensor to notify the server if a specific reading is found. Traditionally, the uploading of code can be done manually for each end device. However, considering a large scale system, this manual task becomes too tedious.

### 2.1.2. Backhaul Links

In order to ensure that these end devices are working as intended, there will be a need to perform testing and debugging, which can only be achieved by extracting the backhaul data. From the data, we can easily determine failed transmissions, successful transmission rates and whether there is any time drifting.

However, the extraction of backhaul data also becomes difficult to achieve due to the bandwidth being possibly congested by uplink transmissions from the large number of end devices. This congestion slows down the rate of backhaul data extraction, delaying the speed of which a fault can be corrected by the system. This problem will definitely be worsened under real-time scenarios with a large number of nodes.

### 2.1.3. Memory Limitations

We can increase the scale of the testbed, but we can't increase the memory size of microcontrollers. Microcontrollers rarely come in memory sizes larger than 10 megabytes. Mathematically, a character is 8 bits. Considering a single event in a scheduler, where each event is represented by "*{3, 1, 8, 1},*" , a single event would take up 13 bytes. If a single event takes places every second, that would mean 60*60*24 events in a day, bringing the number to 86,400 events. Assuming that a microcontroller is responsible for 10 end devices, this number goes up to 864,000. Finally, by multiplying the number of events with the size of a single event, we get 11,232,000 bytes, which is around 11 MB. This number is not even final as the number of devices and events might still increase due to the large-scale nature of the testbed. These calculations show how easy a microcontroller can run out of flash memory space and its incompatibility to be used in a large-scale real-time LoRa testbed.

### 2.1.4. Synchronization Difficulties

In order to deploy real-time schedules across a testbed, time synchronization across all devices is important. This is to ensure that all nodes operate on a shared schedule because the nodes' actions need to be coordinated precisely to avoid any packet collisions and missing deadlines. Being able to precisely activate nodes only when it is required allows for a more energy-efficient setup too. However, this is also hard to achieve due to microcontrollers' innate susceptibility to clock drifting. Microcontrollers' internal clocks are usually powered by an electro-mechanical component, such as a quartz oscillator. The clock drifting of microcontrollers is caused by mechanical and physical reasons near impossible to prevent, such as temperature, aging [15] and inconsistent manufacturing. Over time, these minor time discrepancies accumulate, causing synchronization across each other difficult to achieve. Scaling up the size of the testbeds will further amplify the de-synchronization. Hence, synchronization is infeasible when solely relying on the microcontrollers' internal clocks.

## 2.2. Advantages of a Physical Real-Time LoRa Testbed Over Simulations

Having access to large-scale real-time LoRa testbeds would provide many benefits to the real-time community that simulations cannot bring. Firstly, having a real and physical testbed will allow for the validation of protocols, ensuring that the setups are able to perform according to expectations under accurate hardware and environmental limitations rather than the simplified and easily overlooked constraints in simulations. To be able to test out the setup physically will also help to identify any physical limitations, environmental interferences and any unexpected interactions within the setup.

Secondly, having a physical testbed will allow precise measurement of power consumption of a protocol. Unlike simulations, where estimations or theoretical models might be used for power values, the physical testbed will provide "actual" power values crucial for measuring energy consumption, which would be used later on to optimize the energy consumption of the entire setup.

# 3. Primer

## 3.1. Overview of LoRa Modulation

To assist in better understanding and appreciation of the design principles and logic of the design of the testbed, this section will be discussing the basics on LoRa. Modulation, which refers to converting information into radio waves, is how wireless communication technologies usually transmit data [1]. LoRa modulation works differently due to the use of Chirp Spread Spectrum (CSS). By definition, a chirp is a signal whose frequency changes with time. In LoRa, information is encoded by modulating the frequency of chirps. Each LoRa symbol is made of a series of chirps, and the exact pattern of frequency changes within the chirp determines the data to transmit. An up-chirp represents an increase in frequency while a down-chirp represents a decrease in frequency. CSS, which represents the physical layer modulation of LoRa, provides resistance against interference, allowing for transmission of data across noisy environments and large distances. In this paper, more focus will be put into the parameters – Spreading Factor (SF), Channel (CH) and Bandwidth (BW)[2].

**Spreading Factor** – A setting that will control the chirp rate and thus the speed of data transmission, where a lower spreading factor entails a faster chirp rate. A higher spreading factor also means a longer range, higher Time on Air (ToA), lower data rate, and increased energy consumption [3][4].

**Channel & Frequency -** Frequency refers to the specific radio wave frequency where data will be transmitted while channel refers to one specific frequency. For example, channel_1 will hold 868.10000 while channel_2 will hold 878.10000. In the context of LoRa, there are allocated license free frequency bands allocated to different regions. The choice of frequency is important as too many transmissions on a single channel can cause congestion and high levels of interference.

**Bandwidth** – The bandwidth used will affect the data rate of transmissions. Higher bandwidth leads to a faster rate of transmissions and thus lower ToA. However, a higher bandwidth also means higher noise, and thus lower sensitivity. This lowered sensitivity will prevent weaker signals from being picked up. There is often a need to consider the balance between data transmission speed and sensitivity.

### 3.2. Overview of LoRa Device Communication via SPI Bus

Typically, a LoRa device will require another device to give it instructions, such as a microcontroller. These two items connect via the SPI bus using the pins Slave Select (SS), Master-In-Slave-Out (MISO), Master-Out-Slave-In (MOSI), Serial Clock (SCK) [6][7].

**SS** – Main role is to select the device to do communications with. In this case, it would be the selection of LoRa devices. SS is active low and if multiple SS pins are connected, only one should be active at a time.

**MISO** – Main role is to serve as the main medium for communications done from LoRa device (slave) to the microcontroller (master). The relationship between the master and the slave is such that only the master will request and send data. The slave will not send any data unless instructed to.

**MOSI** – Similar to MISO, but this is for communications done from microcontroller to LoRa device.

**SCK** – Main role is to synchronize data transmission between master and slave devices through the use of clock signals. MISO and MOSI are both synchronized to this clock signal. Due to the use of MISO, MOSI and synchronized communication, SPI is capable of doing full-duplex communication.

Thus, a simple setup between a single LoRa device, which we will consider the SX1262 LoRa radio, can be easily wired up like this:

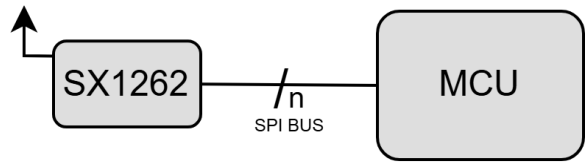

Figure 3.1: Single-Master-Single-Slave Setup Configuration

## 3.3. Single-Master-Multiple-Slave Setup and Transition to Raspberry Pi

In theory, SPI is able to support a Single-Master-Multiple-Slave setup mainly due to the SS mechanism. At any given time, all slaves will share the same lines except for the SS pin. Only one slave's SS pin will be active at a time as the master pulls the desired slave's SS pin low. Meanwhile, the other slave's SS pins will remain high. This way, the communication done between the master and the desired slave is isolated from the rest, preventing any data collisions and complications. Hence, a Single-Master-Multiple-Slave set up can be set up like this:

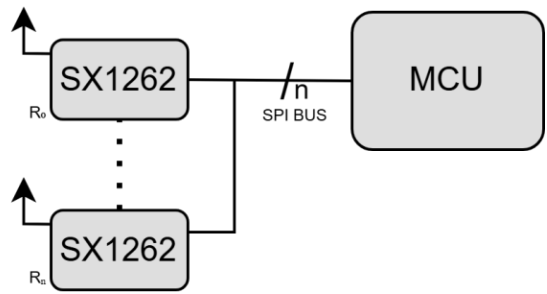

Figure 3.2: Single-Master-Multiple-Slave Setup Configuration (MCU)

However, even though this setup can be easily achieved, it still doesn't contribute to the goal of building a scalable and easy-to-deploy real-time LoRa testbed because the problem of limited memory space of microcontrollers remains. Therefore, to overcome this, we consider the feasibility of using a Raspberry Pi

(RPI) to replace the role of a microcontroller. On a high level of things, a RPI supports the SPI protocol and does not have the same memory constraints as microcontrollers. Popular microcontrollers such as STM32 chips rarely ever have non-volatile memory spaces more than 5MB, while the RPI's memory space depends on the SD card used, with a minimum of 8GB. Based on these 2 factors, a large step would be taken towards our main goal, whilst maintaining a familiar setup:

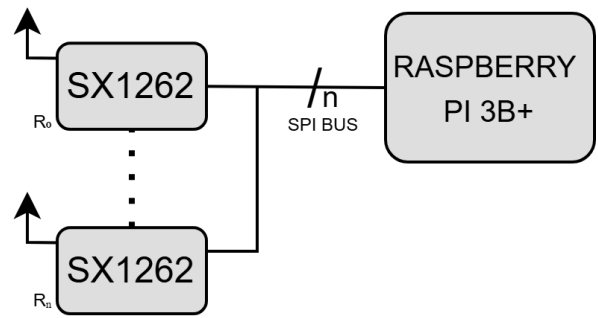

Figure 3.3: Single-Master-Multiple-Slave Setup Configuration (RPI)

# 4. Methodology

## 4.1. Methodology Description

In this section, we will be exploring the feasibility of using the RPI in the aforementioned Single-Master-Multiple-Slave set up and how multiple of these set up can be synchronized together to create a scalable and easy-to-deploy real-time LoRa testbed. Firstly, attempting to use a single RPI to control multiple LoRa radios will be a challenging task because this is not a path well documented and explored.

LoRa and Internet of Things (IoT) devices are conventionally paired together with microcontrollers [8]. The end devices have a unique task they can perform, such as detecting movement, transmitting radio waves, reading temperatures et cetera. However, they do not have the capability to decide when to perform these tasks or what to do with the information, hence the need for a "brain". To fulfil this role, there are options such as microcontrollers, microprocessors, and others. However, microcontrollers fit this role the best because of short development times and lower manufacturing costs [9]. As such, out of many open source LoRa libraries available online, most are designed for microcontroller architectures and for the Arduino IDE.

Regarding the power management for this setup, there might also be concerns regarding insufficient power supply when too many radios are connected to the RPI. There will also be a need to solve the problem of how the multiple radios connected to the RPI can be controlled simultaneously.

## 4.2. LoRa Library for Raspberry Pi

### 4.2.1. Hardware Abstraction Layer

Fortunately, we found that RadioLib [10], a widely used LoRa library, added support for the RPI architecture in 2022, and is still being updated as of today. Utilizing this library, the RPI can manipulate the LoRa radios.

The first main component of RadioLib is the Hardware Abstraction Layer (HAL) [13]. The HAL can be defined as any software that provides a standardized interface between the operating system or application software and the underlying hardware architecture. In other words, the HAL will allow the RPI to be able to instruct the connected LoRa radios to do transmissions via a program.

The very first step of establishing this connection is by gaining control over the RPI pins that are physically connected to the LoRa radios. LGPIO, a library that allows the control of General-Purpose Input Output (GPIO) pins [14], plays a major role in RadioLib's HAL. The use of LGPIO facilitates the configuring and monitoring of the GPIO pins of the RPI. This is done through read and write functions, where each pin can be configured into input/output modes or active-high/active-low.

It is also important to note that the HAL would not be able to work in this setup if not for the SPI0 kernel driver in the Linux kernel. Spidev, an SPI device interface, exposes device files such as "/dev/spidev0.0" which will be crucial in allowing programs to send data through the SPI bus. LGPIO contains SPI-related functions that streamline the process of initializing communication and performing data transfers. By abstracting low-level details, LGPIO allows software and codes to interact with the hardware of the system via the SPI. Once communication via the SPI is initialized, the SPI0 kernel driver will take instructions from the spidev file. As the RPI pins are connected to the LoRa radios, the radios would receive the same instructions. This is how data will be exchanged between the RPI and LoRa radio via the SPI bus and this is how communication between the RPI and LoRa radios become manipulatable through the means of software. The next step would be figuring out what to configure the GPIO pins into.

As mentioned previously, the four major pins of the SPI bus are the SS, MOSI, MISO and SCLK pins. Therefore, manipulating these pins will be foundational to efficient communication between the RPI and the LoRa radios. An example process of the RPI sending data to the LoRa radio would take place like this: Firstly, the SS pin is pulled low, signaling to the LoRa radio that it has been selected for communication. Next, a payload will be prepared in the code to be sent to the LoRa radio via SPI protocol. The MOSI will be manipulated according to the data to be sent while the SCLK pin will be used for synchronization. Afterwards, data transmission is complete, and the SS pin is released back to high. To be able to control these pins properly will ensure data between the master and slave is correctly transmitted and received.

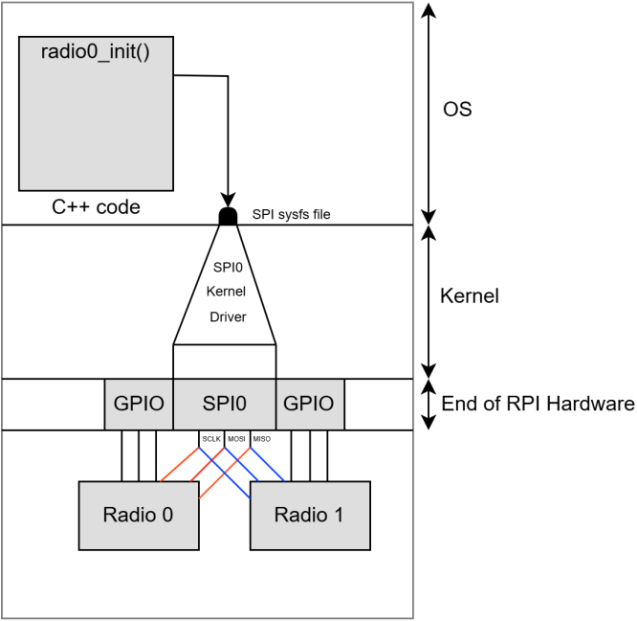

Figure 4.1: Simplified Model Showcasing RPI-Radio Communication via Code

### 4.2.2. Main Code

The second main component of the RadioLib library is the main code that will be run to instruct the LoRa radios to transmit. In the main code, besides the standard instantiating of HAL and radio objects, we bring our focus to the bread and butter of the code – radio.begin() and radio.transmit().

radio.begin() – Main purpose is to initialize the radio and ensure that the radio is ready to send or receive information. The function achieves this by making use of the HAL, radio object instantiations and initializing the radio with key parameters such as frequency, SF, BW and many more.

radio.transmit() – Main purpose is to transmit data via the LoRa radios. This function first runs a standby() function to ensure that the radio halts any ongoing transmissions and is ready for a new transmission. It then checks to ensure that the data packet to be transmitted is of appropriate length and that the timeout timing is properly set accordingly to the Time-On-Air (TOA) parameter, start the transmission via the startTransmit function and enter a loop to wait for either the transmission to complete or timeout. Finally, it records the data rate for performance analysis and finishes the transmission via the finishTransmit function.

```cpp
int16_t SX126x::begin(uint8_t cr, uint8_t syncWord, uint16_t preambleLength, float tcxoVoltage, bool useRegulatorLDO) {
  // BW in kHz and SF are required in order to calculate LDRO for setModulationParams
  // set the defaults, this will get overwritten later anyway
  this->bandwidthKhz = 500.0;
  this->spreadingFactor = 9;

  // initialize configuration variables (will be overwritten during public settings configuration)
  this->bandwidth = RADIOLIB_SX126X_LORA_BW_500_0;  // initialized to 500 kHz, since lower values will interfere with LLCC68
  this->codingRate = RADIOLIB_SX126X_LORA_CR_4_7;
  this->ldrOptimize = 0x00;
  this->crcTypeLoRa = RADIOLIB_SX126X_LORA_CRC_ON;
  this->preambleLengthLoRa = preambleLength;
  this->tcxoDelay = 0;
  this->headerType = RADIOLIB_SX126X_LORA_HEADER_EXPLICIT;
  this->implicitLen = 0xFF;

  // set module properties and perform initial setup
  int16_t state = this->modSetup(tcxoVoltage, useRegulatorLDO, RADIOLIB_SX126X_PACKET_TYPE_LORA);
  RADIOLIB_ASSERT(state);

  // configure publicly accessible settings
  state = setCodingRate(cr);
  RADIOLIB_ASSERT(state);

  state = setSyncWord(syncWord);
  RADIOLIB_ASSERT(state);

  state = setPreambleLength(preambleLength);
  RADIOLIB_ASSERT(state);

  // set publicly accessible settings that are not a part of begin method
  state = setCurrentLimit(60.0);
  RADIOLIB_ASSERT(state);

  state = setDio2AsRfSwitch(true);
  RADIOLIB_ASSERT(state);

  state = setCRC(2);
  RADIOLIB_ASSERT(state);

  state = invertIQ(false);
  RADIOLIB_ASSERT(state);

  return(state);
}
```

Figure 4.2: Snippet of Function radio.begin()

```cpp
int16_t SX126x::transmit(const uint8_t* data, size_t len, uint8_t addr)
{ // set mode to standby
  int16_t state = standby();
  RADIOLIB_ASSERT(state);

  // check packet length
  if(len > RADIOLIB_SX126X_MAX_PACKET_LENGTH) {
    return(RADIOLIB_ERR_PACKET_TOO_LONG);
  }

  // calculate timeout in ms (5ms + 500 % of expected time-on-air)
  RadioLibTime_t timeout = 5 + (getTimeOnAir(len) * 5) / 1000;
  RADIOLIB_DEBUG_BASIC_PRINTLN("Timeout in %lu ms", timeout);

  // start transmission
  state = startTransmit(data, len, addr);
  RADIOLIB_ASSERT(state);

  // wait for packet transmission or timeout
  uint8_t modem = getPacketType();
  RadioLibTime_t start = this->mod->hal->millis();
  while(true) {
    // yield for multi-threaded platforms
    this->mod->hal->yield();

    // check timeout
    if(this->mod->hal->millis() - start > timeout) {
      finishTransmit();
      return(RADIOLIB_ERR_TX_TIMEOUT);
    }

    // poll the interrupt pin
    if(this->mod->hal->digitalRead(this->mod->getIrq())) {
      // in LoRa or GFSK, only Tx done interrupt is enabled
      if(modem != RADIOLIB_SX126X_PACKET_TYPE_LR_FHSS) {
        break;
      }

      // in LR-FHSS, IRQ signals both Tx done as frequency hop request
      if(this->getIrqFlags() & RADIOLIB_SX126X_IRQ_TX_DONE) {
        break;
      } else {
        // handle frequency hop
        this->setLRFHSSHop(this->lrFhssHopNum % 16);
        clearIrqStatus();
      }
    }
  }

  // update data rate
  RadioLibTime_t elapsed = this->mod->hal->millis() - start;
  this->dataRateMeasured = (len*8.0)/((float)elapsed/1000.0);

  return(finishTransmit());
}
```

Figure 4.3: Snippet of Function radio.transmit()

With this, we will have a working system that can support the RPI architecture in acting as the "brain" of the LoRa radios. The only thing left to do will be to make changes to the main code that will accommodate the Single-Master-Multiple-Slave setup. For each LoRa radio connected, the HAL and Radio object will have to be instantiated, and initialized via radio.begin(). Afterwards, parameters need to be set correctly to ensure the transmissions are successful, such as frequency, SF, synchronization word, and output power. The parameters in this segment can be changed according to the setup's needs. Lastly, in the end product, we will make use of the Time Division Multiple Access (TDMA) protocol, and a scheduler will be used to instruct the radios when transmissions should be made.

```cpp
int main(int argc, char** argv) {
  // Initialize radios
  PiHal* hal = new PiHal(0);
  PiHal* hal2 = new PiHal(0);
  PiHal* hal3 = new PiHal(0);

  // Create radio instances
  SX1262 radio1 = new Module(hal, 12, 27, 25, 19);
  SX1262 radio2 = new Module(hal2, 22, 23, 24, 20);
  SX1262 radio3 = new Module(hal3, 5, 6, 13, 26);
```

Figure 4.4: Snippet of Main Code – Instantiating Objects

```cpp
printf("[SX1262] Initializing Radio 1 ... ");
int state = radio1.begin();
if (state != RADIOLIB_ERR_NONE) {
  printf("failed, code %d\n", state);
  return 1;
}
printf("success!\n");

printf("[SX1262] Initializing Radio 2 ... ");
state = radio2.begin();
if (state != RADIOLIB_ERR_NONE) {
  printf("failed, code %d\n", state);
  return 1;
}
printf("success!\n");

printf("[SX1262] Initializing Radio 3 ... ");
state = radio3.begin();
if (state != RADIOLIB_ERR_NONE) {
  printf("failed, code %d\n", state);
  return 1;
}
printf("success!\n");
```

Figure 4.5: Snippet of Main Code – Systematically Initializing Radios

```
radio1.setFrequency(868.1);
radio2.setFrequency(868.1);
radio3.setFrequency(868.1);

radio1.setSpreadingFactor(12);
radio2.setSpreadingFactor(12);
radio3.setSpreadingFactor(12);

radio1.setSyncWord(0x34, 0x44);
radio2.setSyncWord(0x34, 0x44);
radio3.setSyncWord(0x34, 0x44);

radio1.setOutputPower(0);
radio2.setOutputPower(0);
radio3.setOutputPower(0);
```

Figure 4.6: Snippet of Main Code – Setting Parameters

```
int count = 0;
for (;;) {
  // Send packet from Radio 1
  printf("[SX1262] Transmitting packet from Radio 1 ... ");
  char str[64];
  sprintf(str, "Hello from Radio 1! #%d", count++);
  state = radio1.transmit(str);
  if (state == RADIOLIB_ERR_NONE) {
    printf("success!\n");
  } else {
    printf("failed, code %d\n", state);
  }
```

Figure 4.7: Snippet of Main Code – Transmitting from Single Radio

## 4.3. Power Concerns

Regarding the power management for this setup, the SX1262 LoRa radios will be standardized as EByte's E22-900M30S modules [11], along with a Raspberry Pi 3B+ [12]. Since the SPI bus is a shared resource, when testing out transmissions with multiple radios under a multithreaded program, there will be a need for a mutex to ensure no unexpected behaviors. As such, only one radio will be doing transmissions. The other radios will remain in idle mode as they wait for the mutex to unlock. Therefore, we are only interested in these values:

| Purpose | Current(A) |
| --- | --- |
| RPI - Current Supplied | 2.5 |
| RPI - Bare-board Active Current | 0.5 |
| SX1262 - Max Idle Current | 0.000000005 (Negligible) |
| SX1262 - Max Transmit Current | 0.7 |

From this table, it can be calculated that the RPI will have 2.0A that can be used by the SX1262 radios. Since only one radio will be transmitting at most, there will be ample power supply left, leaving no concerns for a case of power shortage.

## 4.4. Simultaneous Control

As mentioned earlier, doing true simultaneous transmissions over the LoRa radios will not be possible. The reason for being unable to have truly simultaneous control is this – the SPI bus is a single shared bus. At any given time, only one SS pin should be activated and instructed. Should more than one SS pin be activated, the instruction transmitted from the RPI to the designated LoRa radio will also be transmitted to the other LoRa radios, leading to potentially unwanted parameter changes or transmissions. Similarly, if the RPI is expecting data from Radio 1 via the MISO, the data could be corrupted if Radio 2 unknowingly sends data via the same MISO. Hence, to prevent this, there is a need for a mutex to lock any SPI instructions and in exchange, the program will not be truly simultaneous.

Although true parallel transmissions will not be possible, we can still create a program that is nearer to being true parallel. The structure of the program can be changed into a multithreaded one. This way, each radio will have their own thread to handle the transmission and setting of parameters. Multithreading can be applied easily onto the main code from RadioLib by using specific libraries from C++. Additionally, the mutex to prevent any complications can be placed before the setting of any unique parameters (SF and CH) and transmit instructions for a radio and be unlocked after.

## 4.5. Synchronizing RPIs

With the building blocks of the testbed established, the next step would be to attempt to synchronize these building blocks to ensure that schedules are followed in a correct and consistent manner by each block. For example, the 5th second of RPI_1 will have to be the 5th second of every other RPI in the testbed. This way, unexpected outcomes such as overlapping transmissions, packet collision and out of sync timestamps can be avoided.

For this, a Network Time Protocol (NTP) will be utilized [16]. To put it simply, an NTP works by having the RPI ask a trusted time server for the "correct" time and adjust to it accordingly. This way, it doesn't matter how many RPIs will be used in the testbed, if the time server contacted is the same, the RPIs will all stay under the same synchronized time with an acceptable level of accuracy.

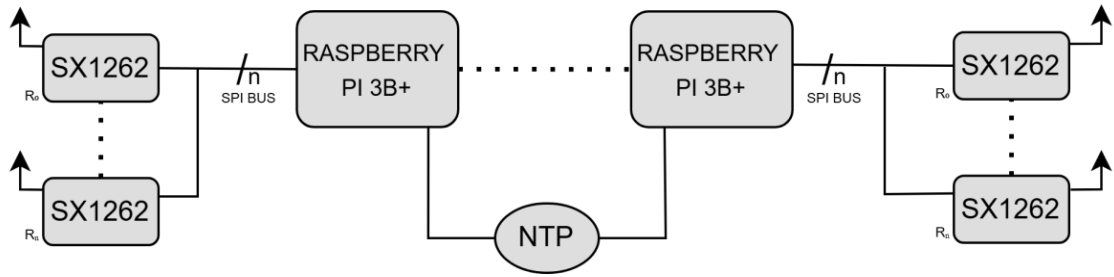

Figure 4.8: Complete Testbed Configuration

# 5. Experiments

## 5.1. Experiment Goals

The goal of this experiment is to evaluate the feasibility and performance of a small component of the testbed - one RPI connected to three LoRa radios. This experiment will serve to verify that the setup works as intended and the RPI is able to instruct the three radios to transmit in a multithreaded program, improving the setup's portability and organization so that it looks closer to an end-product, and finally evaluating the performance of the setup by testing simple scheduling protocols.

## 5.2. Preliminary Testing

The most basic configuration that can be built is to make sure the RPI is able to initialize and instruct more than one radio to transmit. The very first step to achieving this would be to connect the RPI and radios physically.

### 5.2.1. Hardware Configurations

The SX1262 Radios that will be worked on come with already-labeled pads. The more relevant pins includes:

**SPI Interface** - NSS, MOSI, MISO, and SCLK. These pins work in the same way as previously discussed, but there is a slight difference in naming.

**DIO1** - Used as the digital input/output pin.

**NRST** - Used as the reset signal.

**BUSY** - Used as the busy indicator.

**VCC** - For power.

**GND** - For grounding.

The MOSI, MISO and SCLK pins have to be connected to designated pins on the RPI, as well as the VCC and GND pins, while the other pins just have to be connected to any GPIO pins. For this case, we chose pins 12, 19, 25 & 27.

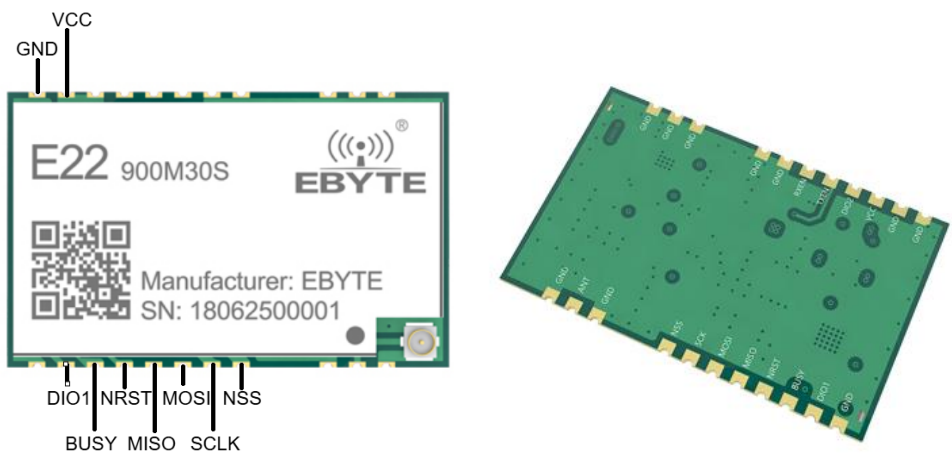

Figure 5.1 & 5.2: Front & Back of Used SX1262 Radio [11]

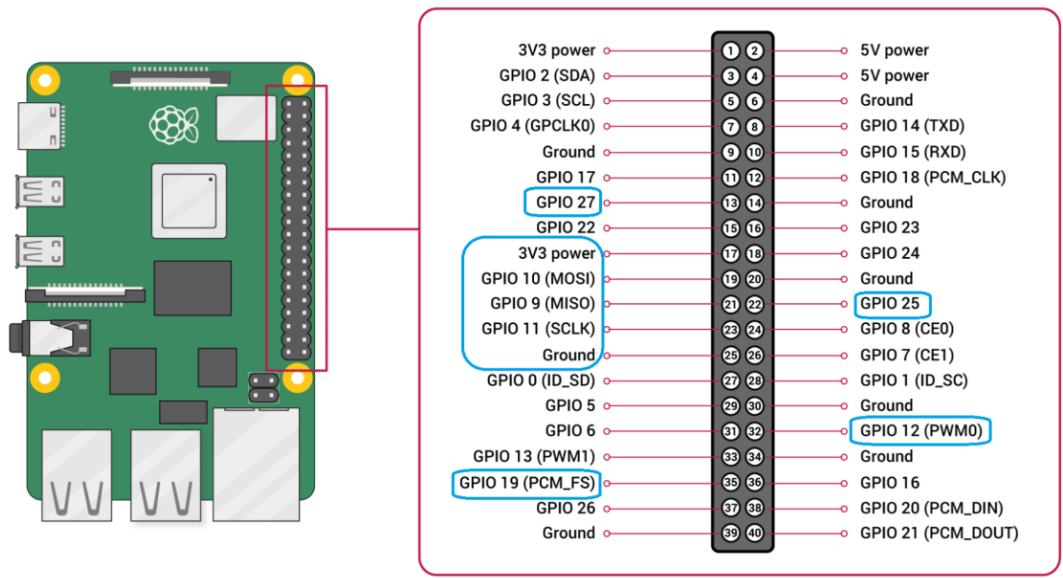

Figure 5.3: Pins Layout of Raspberry Pi [12]

### 5.2.2. Writing the Program

With the help of RadioLib's example code, only small changes had to be made to the code [10]. Since the code was written in C++, there is a need to correctly compile the program and include correct flags such as "*-llgpio*" and "*-lRadioLib*". By running the code with correct parameters, pin configurations, it can be observed that the transmission and connection was a success.

```c
#include <RadioLib/RadioLib.h>
#include "PiHal.h"

PiHal* hal = new PiHal(0);
SX1262 radio = new Module(hal, 12, 25, 27, 19);

// the entry point for the program
int main(int argc, char** argv) {
  // initialize just like with Arduino
  printf("[SX1261] Initializing ... ");
  int state = radio.begin();
  if (state != RADIOLIB_ERR_NONE) {
    printf("failed, code %d\n", state);
    return(1);
  }
  printf("success!\n");

  radio.setFrequency(868.1);
  radio.setSpreadingFactor(12);
  radio.setSyncWord(0x34, 0x44);
  radio.setOutputPower(0);

  // loop forever
  int count = 0;
  for(;;) {
    // send a packet
    printf("[SX1261] Transmitting packet ... ");
    char str[64];
    sprintf(str, "Packet #%d from Radio 1", count++);
    state = radio.transmit(str);
    if(state == RADIOLIB_ERR_NONE) {
      // the packet was successfully transmitted
      printf("success!\n");

      // wait for a second before transmitting again
      hal->delay(1000);

    } else {
      printf("failed, code %d\n", state);

    }

  }
  return(0);
}
```

Figure 5.4 & 5.5: Snippet of Code Used

```
pi2@raspberrypi:/home/pi/Desktop/RadioLib/examples/NonArduino/Raspberry $ sudo ./main
[SX1261] Initializing ... success!
[SX1261] Transmitting packet ... success!
[SX1261] Transmitting packet ... success!
[SX1261] Transmitting packet ... success!
[SX1261] Transmitting packet ... success!
[SX1261] Transmitting packet ... success!
[SX1261] Transmitting packet ... success!
[SX1261] Transmitting packet ... success!
[SX1261] Transmitting packet ... success!
```

Figure 5.6: Successful Transmission Logs (From Radio's Terminal)

```
Decoded LoRa Data Log
=====================
Received message from ('127.0.0.1', 48967):
Full Decoded JSON:
▫!S UZ
▫▫{"rxpk":[{"tmst":4162173235,"chan":1,"rfch":1,"freq":868.100000,"stat":1,"modu":"LORA","datr":"SF12BW125","codr":"4/7","lsnr":7.5,"rssi":-84,"size":21,
"data":"UGFja2V0IDAgZnJvbSBSYWRpbyAx"}]}
Decoded Base64 Data: Packet 0 from Radio 1

Received message from ('127.0.0.1', 48967):
Full Decoded JSON:
▫~ UZ
▫▫{"rxpk":[{"tmst":4162260755,"chan":1,"rfch":0,"freq":868.100000,"stat":1,"modu":"LORA","datr":"SF12BW125","codr":"4/7","lsnr":10.0,"rssi":-82,"size":21
,"data":"UGFja2V0IDAgZnJvbSBSYWRpbyAx"}]}
Decoded Base64 Data: Packet 1 from Radio 1
```

Figure 5.7: Successful Transmission Logs (Formatted Logs from Gateway)

## 5.3. Testing Multiple Radios

Following up from the previous setup, the simplest way to verify that multiple radios can work with a single RPI is to connect exactly one more radio to the RPI and instantiate another radio in the code. In order to connect a second radio, the connections are mostly the same with the first radio, except a few points to take note: 1) The radios should share the same SPI pins, except for the NSS pin. The NSS pin should be unique because the communications made should be exclusive to one radio at any given time. 2) The connections made for power and ground can be shared. 3) The connections made for NRST, BUSY, DIO1 have to also be uniquely connected to other GPIO pins. In the program, there will be a need to instantiate another HAL object and use it to initialize the second radio.

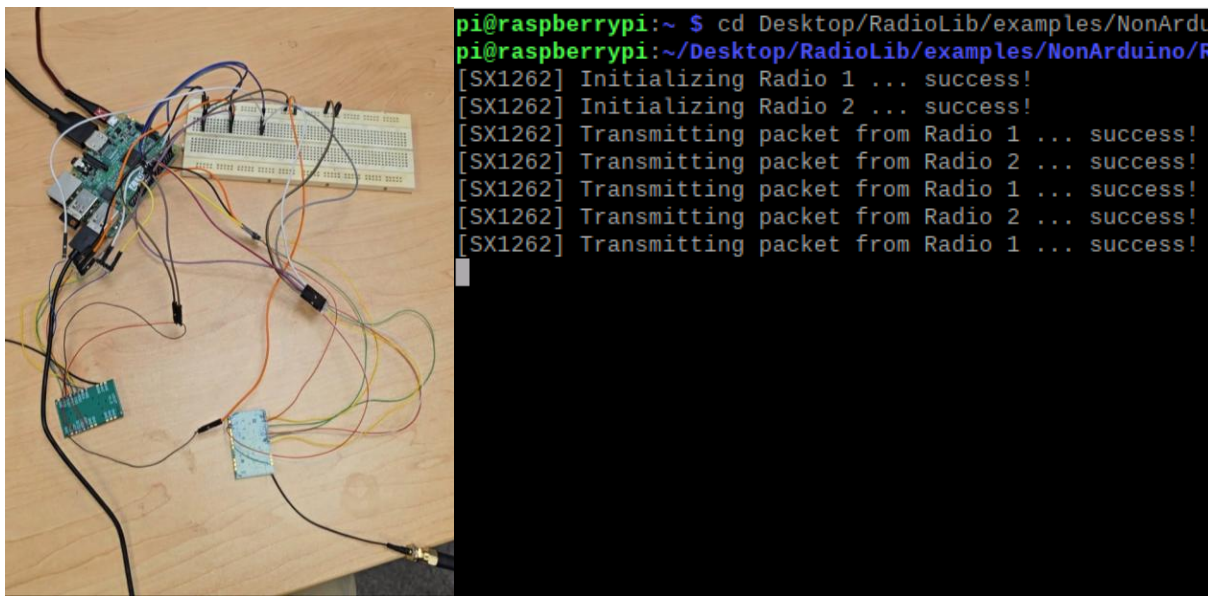

Figure 5.8 & 5.9: Setup Used & Logs Observed

## 5.4. Creating the Final Setup

After verifying that the setup works with more than one radio, the next step was to add a third radio, which is also the end-setup that would be worked with. However, as seen from the initial setup, the wirings can get messy and hard to move around, hence a need to make it neater and more portable. With the help of a flat cable extension for the RPI, we managed to turn the setup into this:

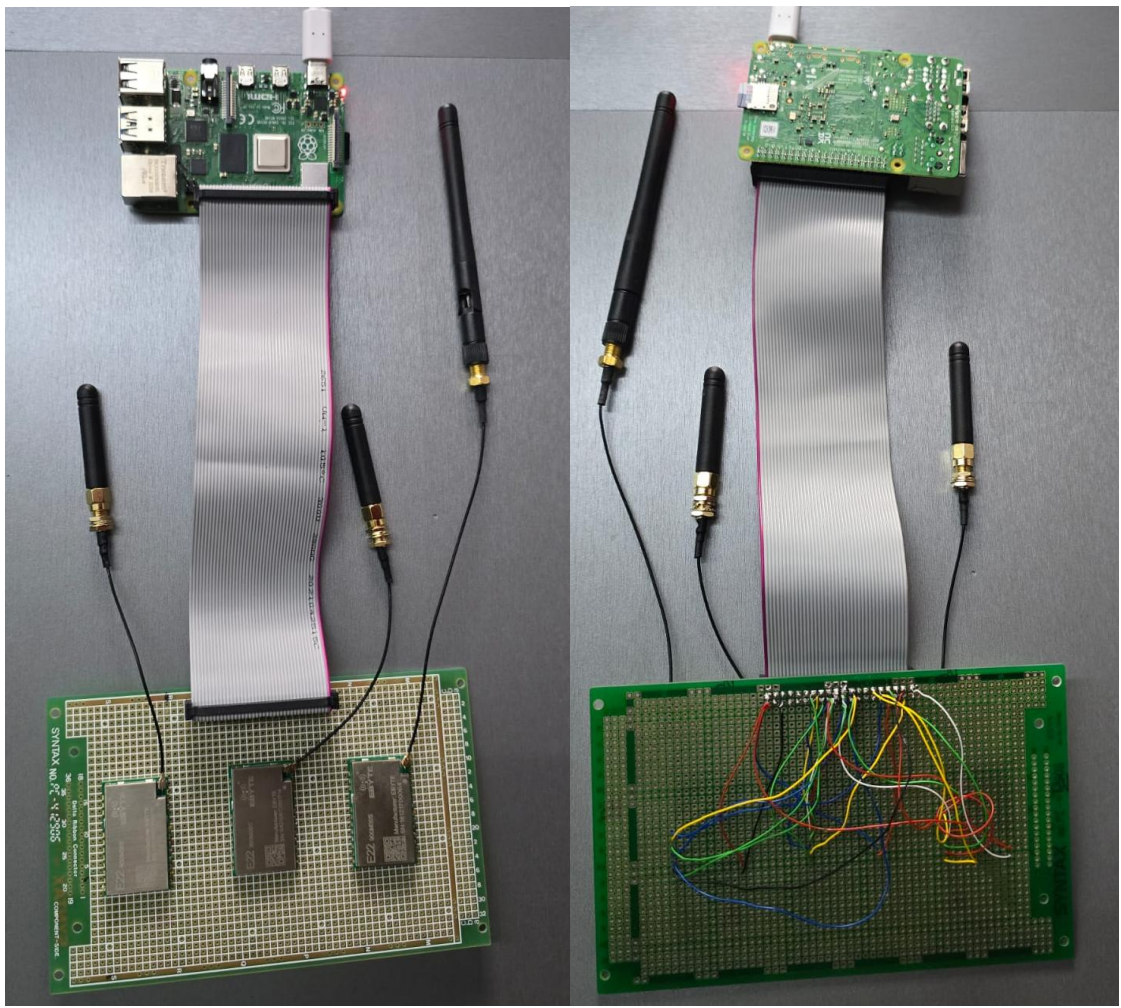

Figure 5.10 & 5.11: Top & Bottom of Final Setup

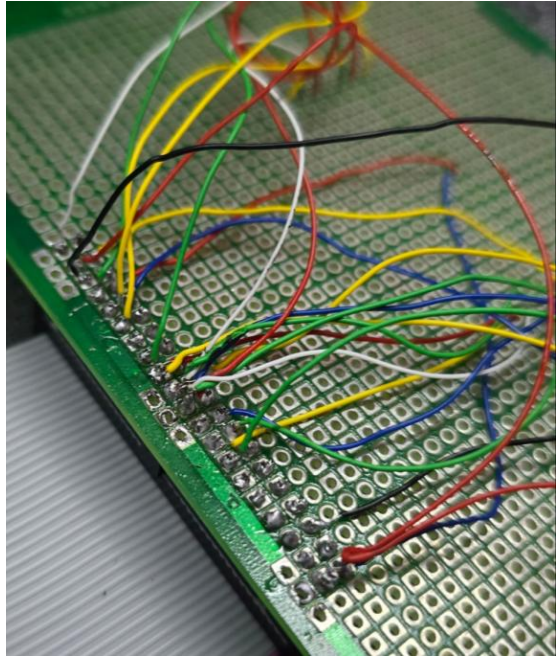

Figure 5.12: Wire Configuration

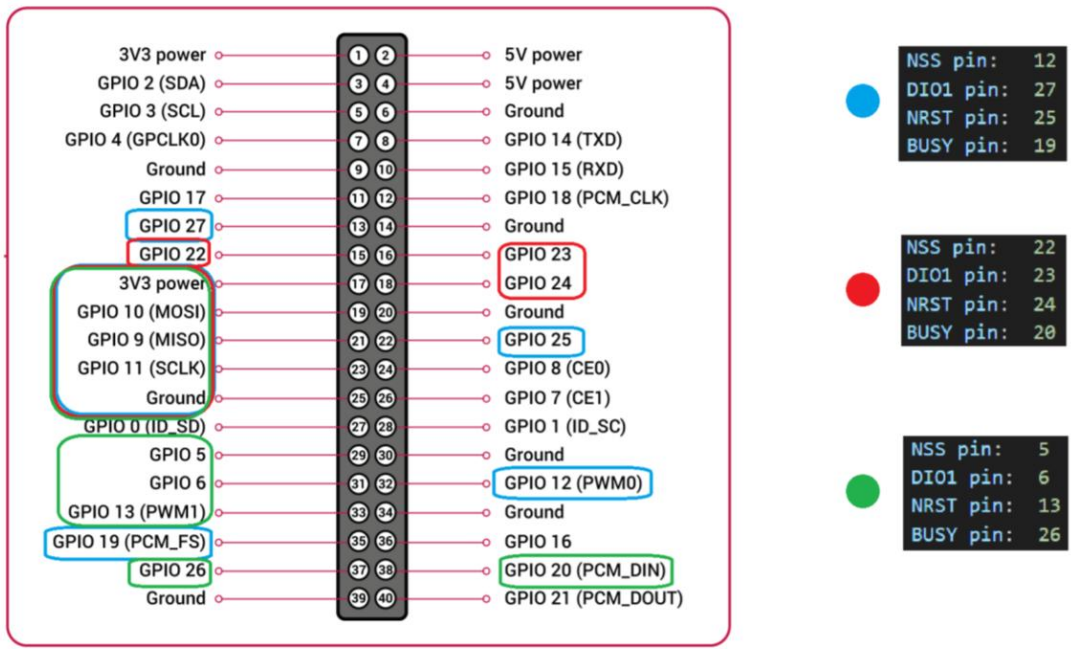

Figure 5.13: Final Configuration for Connections

As mentioned previously, the program will be a multithreaded one so that the setup can get the radios to transmit as simultaneously as possible. How the code works is still roughly similar to when one radio was tested, except now, instead of each radio transmitting every second, schedulers will be used to control transmissions.

## 5.5. Performance Evaluation

### 5.5.1. Schedules

As mentioned earlier, schedules will be used to dictate when the radios are supposed to transmit. In our experiments, the schedule follows a certain formatting. Our schedules are assigned to each radio in the code, where each assignment is done with an array of arrays called "*schedulesX[ ]*". Within it, the sub-arrays will contain four items: the node (radio) ID, when to transmit, at which spreading factor and which channel. An example schedule can be "*{0, 3, 7, 1}*", which would mean node 0 (Radio 1) will transmit at the third second, with spreading factor 7 on channel 1. The channel will be initialized as an array of frequencies in my code, hence the number in the schedule will correspond to the index of the frequency. By writing a python script that generates a list of events, schedules can be created easily for testing.

### 5.5.2. Test Cases

In order to assess how well the setup will work, the performance per test case will be evaluated using a few important metrics:

**Received Signal Strength Indicator (RSSI)** - Measurement of power present in a received radio signal. The measurement is represented in negative values, where the closer the value to 0, the better. RSSI can be affected by a few factors such as hardware, distance, interference and environment [17].

**Signal-to-Noise Ratio (SNR)** - The higher the value, the clearer the signal and the easier to detect.

**Packet Loss Ratio (PLR)** - The higher the value, the more undesirable the communication is.

**Packet Received Ratio (PRR)** - It is the inverse of PLR and the higher the value, the more reliable the communication is.

**Latency** - In the context of real-time, latency is a very important criteria to meet. Latency can be calculated by taking (time of receiving at gateway - time of transmitting from RPI).

**Jitter** - Jitter refers to the consistency in transmission timings. By comparing the delays of each transmission timing, the jitter can be derived by measuring any fluctuations.

In our experiments, the transmissions have very small transmissions periods with millisecond level accuracy. Hence, latency and jitter become important metrics for us to measure. To evaluate these metrics, some specific experiments were conducted and their implications on the system were also discussed.

### 5.5.3. Execution Workflow

For each test case, the experimental procedure goes like this. Firstly, using the python script, schedules are generated for each radio. Manually, the schedules are added into the main code which starts the transmissions. After compiling, the gateway is ensured to be listening and logging, before running the main code. As the main code finishes running, the logs also stop collecting. This same cycle is repeated three times so that an average value can be calculated from the logs later on. Afterwards, the logs are extracted from the gateway and processed by using a python script to extract important and relevant values.

### 5.5.4. Effects of Varying Spread Factor Value

By generating schedules that keep the same channels while varying SF values, the effects of varying SF values can be observed. In this test case, all radios transmitted on the frequency band of 867.300000, while each radio transmitted on SF 7, 8 and 9 respectively.

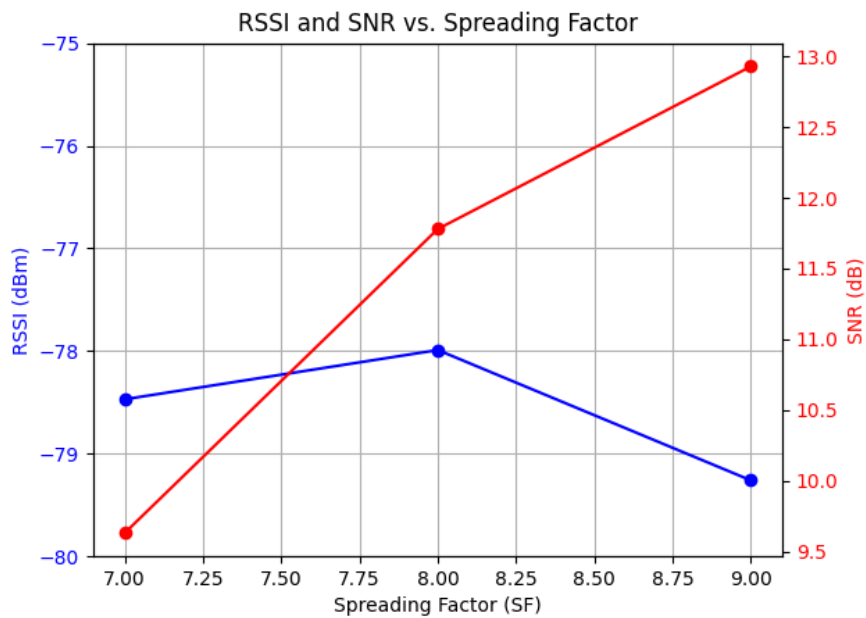

Figure 5.14: RSSI & SNR Across Different SFs

As expected, increasing the SF will cause transmissions to have a higher ToA, thus making the signals more robust to noise, resulting in increasing SNR. For the RSSI, since the setup was not moved at all with respect to the gateway, the RSSI observed did not have any big changes.

### 5.5.5. Effects of Varying Channels

In this experiment, the radios are all configured to transmit on SF 7, with the channels varying between 867.200000 to 868.400000. This experiment serves to show performance of the radios on different frequency bands.

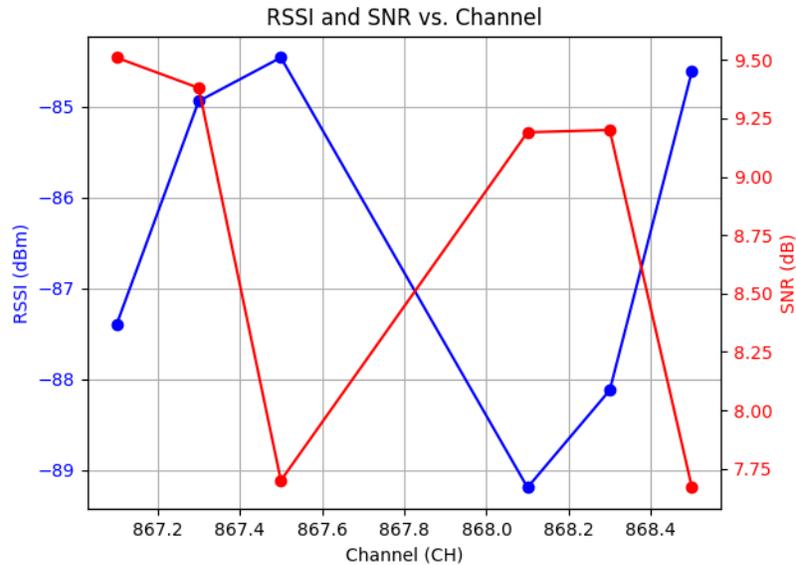

Figure 5.15: RSSI & SNR Across Different CHs

From the findings, it can be seen that varying the channels doesn't really create large differences in terms of RSSI and SNR. This is also to be expected as varying the channels should only have a noticeable impact if the channels used are comparatively busy.

### 5.5.6. TDMA vs CSMA

This experiment's main purpose is to evaluate the performance of our setup, which uses TDMA, against another commonly used medium access protocol known as Carrier Sense Multiple Access (CSMA). The main difference between the two is that TDMA allocates specific time slots for each radio to transmit through the use of schedules, allowing for collision-free communication, while CSMA allows radios to transmit by first checking whether a channel is idle, and if not, it will do channel hopping till an idle channel is found. Initially, the performance of both protocols was to be assessed by comparing the packet loss ratio (PLR), which will give a sense of the efficiency and reliability in terms of packet delivery.

However, due to the absence of true parallel transmissions, CSMA transmissions are not exactly doable. By default, when a channel is in use by the first radio, the second radio is supposed to be doing collision avoidance detection and transmitting on the next available channel. In this case, since the first radio was holding onto the mutex lock, the second radio cannot transmit. The second radio waits for the first radio, then does its back offs, and then transmits. Thus, a fairer comparison will be done between the two MAC's energy consumption.

In TDMA, the energy consumption that we are interested in is the energy used to transmit packets throughout the entire transmission period. From the datasheet of the radio, we can get the communication level/voltage (V), transmitting current (A) and the total time spent transmitting (s), which will give us the energy consumption when multiplied together. In CSMA, we are also interested in the aforementioned values, but there is an additional energy used for doing collision avoidance detection and backoffs which are regarded as energy overheads. In each backoff, the energy overhead is estimated as 0.356 mJ [21]. Due to the limitations in true parallel transmission capabilities, the energy used in transmitting the actual packets will be the same in both, with the only difference being the energy overhead.

Hence, in our experiment, where 206 packets were transmitted, the energy overhead from CSMA can be calculated as 0.356 mJ * 206, giving us 73.336 mJ. Therefore, it can be said that for our current setup, TDMA is more energy efficient due to the lack of energy overheads, which can be calculated by 0.356 mJ * (no. of packets). Being energy efficient is very important under real-time scenarios because as the sensors are expected to transmit very often, the energy usage can get very high, especially in large-scale.

### 5.5.7. Aperiodic vs Periodic

As our setup uses a scheduler, it is important to find out the value of using fixed time periods. Thus, this experiment will aim to compare the performance of aperiodic transmissions vs periodic transmissions, by considering the PLR of these two transmissions. The reason why aperiodic should be considered as a contender is because aperiodic transmissions are useful in cases where data to be sent is irregular and unpredictable. However, the main trade off is that they can lead to more collisions.

The experiment takes place under the conditions of the same number of packets, same SFs and different channels. The main difference is the periodic and aperiodic nature of the schedules. By calculating, it can be found that the aperiodic setup has a mean PLR of 0.30952 while the periodic setup has a mean PLR of 0.24058. As expected, when the packets are not scheduled to be transmitted in an organized manner, the chances of collisions are highly increased.

### 5.5.8. Effects of Using Same Channels & SF

So far, we have found out the effects of varying the channel and varying the SF. Naturally, there would also be a need to find out the effect of not varying either. In this experiment, we will be keeping notes of the PLR, because we expect the PLR of the setup with the same channel and SFs to be the worst, due to the increased rate of packet collisions.

| Setup | PLR |
| --- | --- |
| Same CH, Different SF | 0 |
| Different CH, Same SF | 0.24057971014 |
| Same CH, Same SF | 0.0119047619 |

As observed from the table, it can be seen that Same CH-Same SF actually outperforms Different CH-Same SF. The results can be owed to 2 reasons. Firstly, because of the lack of true parallel transmissions, having transmissions in the same channel and SF doesn't actually mean more collisions. This is because at the very core, only one packet is transmitted. Secondly, there was some anomaly throughout the experiment linked to varying channels, which will be discussed at a later part.

### 5.5.9. Milliseconds Accuracy in Transmission Delay Jitter

In the interest of achieving real-time latency, there is a need for the packets to be delivered on time. In this experiment, the delay jitter of the transmissions will be studied. Delay jitter is important because it evaluates the consistency of the system in terms of transmitting on time. In this experiment, the focus is on the transmission period. By setting a period to as low as possible for the SF of 7, the experiment becomes an edge-case scenario with a single radio.

```
pi2@raspberrypi:~/Desktop/multisched $ sudo ./main
[SX1262] Initializing Radio 1 ... success!
[SX1262] Initializing Radio 2 ... success!
[SX1262] Initializing Radio 3 ... success!
[SX1262-1] Transmitting packet ...
[SX1262-1] success!
[SX1262-1] Elapsed time: 181 ms
[SX1262-1] Calculated delay: -1 ms
[SX1262-1] Transmitting packet ...
[SX1262-1] success!
[SX1262-1] Elapsed time: 269 ms
[SX1262-1] Calculated delay: 1 ms
[SX1262-1] Transmitting packet ...
[SX1262-1] success!
[SX1262-1] Elapsed time: 357 ms
[SX1262-1] Calculated delay: 3 ms
[SX1262-1] Transmitting packet ...
[SX1262-1] success!
[SX1262-1] Elapsed time: 448 ms
[SX1262-1] Calculated delay: 2 ms
[SX1262-1] Transmitting packet ...
[SX1262-1] success!
[SX1262-1] Elapsed time: 537 ms
[SX1262-1] Calculated delay: 3 ms
[SX1262-1] Transmitting packet ...
[SX1262-1] success!
[SX1262-1] Elapsed time: 628 ms
```

Figure 5.17: Logs Showing Calculated Delay

In the logs observed, "Calculated delay" refers to the amount of time to wait in milliseconds, before having to transmit the next packet. Throughout the logs, the value of the calculated delay deviates between 3ms to -2ms. This shows that the setup is able to have milliseconds accurate consistency in the delay jitter of transmissions even at edge cases. This further validated that the architecture of the RPI is able to support the requirements of a real-time LoRa testbed.

### 5.5.10. Finding Minimum Transmission Period

In this experiment, we define minimum transmissions period as the minimum period that should be set for the specific SF, such that excessive missing of deadlines and accumulation of delays will not happen. In other words, when trying to set the smallest period for the specific SF, the period should never be smaller than this.

When considering the transmission period, there is a relationship between the SF and ToA to be taken into account. As understood, a larger SF results in longer ToA. If the scheduler sets periods that are impossible to follow given the SF, a delay is bound to happen, and deadlines might not be met. With reference to a designer guide from Semtech [18] for the SX127X series, there is an online tool which uses formulas from the guide to create a calculator for the ToA of LoRaWAN transmissions [19]. Using the calculator's timings as a starting point, we conducted experiments using those values as the periods, to find what conditions the setup needs to have milliseconds accuracy in transmissions.

For the experiment carried out earlier in 5.5.8 with SF7 and just one radio, the optimal period associated with SF7 was found to be around 90ms. However, in a scenario where all three radios had to transmit at the same time, the optimal period was then found to be around 270ms, three times that of the previous value:

```
pi2@raspberrypi:~/Desktop/multisched $ sudo ./main [SX1262-1] Transmitting packet ...
[SX1262] Initializing Radio 1 ... success!    [SX1262-1] success!
[SX1262] Initializing Radio 2 ... success!    [SX1262-1] Elapsed time: 359 ms
[SX1262] Initializing Radio 3 ... success!    [SX1262-1] Calculated delay: 181 ms
[SX1262-1] Transmitting packet ...            [SX1262-2] Transmitting packet ...
[SX1262-1] success!                           [SX1262-2] success!
[SX1262-1] Elapsed time: 181 ms               [SX1262-2] Elapsed time: 450 ms
[SX1262-1] Calculated delay: -1 ms            [SX1262-2] Calculated delay: 90 ms
[SX1262-1] Transmitting packet ...            [SX1262-3] Transmitting packet ...
[SX1262-1] success!                           [SX1262-3] success!
[SX1262-1] Elapsed time: 269 ms               [SX1262-3] Elapsed time: 537 ms
[SX1262-1] Calculated delay: 1 ms             [SX1262-3] Calculated delay: 3 ms
[SX1262-1] Transmitting packet ...            [SX1262-2] Transmitting packet ...
[SX1262-1] success!                           [SX1262-2] success!
[SX1262-1] Elapsed time: 357 ms               [SX1262-2] Elapsed time: 629 ms
[SX1262-1] Calculated delay: 3 ms             [SX1262-2] Calculated delay: 181 ms
[SX1262-1] Transmitting packet ...            [SX1262-1] Transmitting packet ...
[SX1262-1] success!                           [SX1262-1] success!
[SX1262-1] Elapsed time: 448 ms               [SX1262-1] Elapsed time: 716 ms
[SX1262-1] Calculated delay: 2 ms             [SX1262-1] Calculated delay: 94 ms
[SX1262-1] Transmitting packet ...            [SX1262-3] Transmitting packet ...
[SX1262-1] success!                           [SX1262-3] success!
[SX1262-1] Elapsed time: 537 ms               [SX1262-3] Elapsed time: 803 ms
[SX1262-1] Calculated delay: 3 ms             [SX1262-3] Calculated delay: 7 ms
[SX1262-1] Transmitting packet ...            [SX1262-2] Transmitting packet ...
[SX1262-1] success!                           [SX1262-2] success!
[SX1262-1] Elapsed time: 628 ms
```

Figure 5.18: Comparison Between 1-Radio & 3-Radio Setup

This is because the setup we used is not able to have true parallel transmissions and thus, the minimum transmission period to consistently meet transmission deadlines differs by the number of radios. Following this pattern, the minimum transmission periods for each SF can be found.

### 5.5.11. PRR Across Different Test Cases

After successfully conducting several different test cases, the same set of results are used to observe how the PRR of each test case fares. The test cases that will be used for this experiment are varying SF, varying CH, varying both and varying neither. When the SF is to be unchanged, SF7 was used, while the frequency of 867.300 is used when CH is to be unchanged.

| PRR Table (Without deadlines) | Varying CH | Same CH |
|---|---|---|
| Varying SF | 0.733333333333333 | 0.992857142857142 |
| Same SF | 0.771428571428571 | 0.954761904761904 |

From the results, it can be seen that the best performing combination is varying SF and having the same CH. However, based on previous test cases, there is some anomalous data to be picked on here that will be discussed in the later parts.

Following up on this experiment, an additional condition of meeting deadlines is added. To recall, the main criterion of real-time communication is to meet deadlines, which was the goal of this experiment.

| PRR (With Deadlines) | Varying CH | Same CH |
|---|---|---|
| Varying SF | 0.6666666666666666 | 0.9380952380952381 |
| Same SF | 0.6904761904761905 | 0.9095238095238095 |

As expected, the PRR drops from the previous table. To give some background, LoRa sensors used in real-time context, are able to collect data continuously. However, is it not a common application to have the sensors transmit data continuously. Therefore, in this experiment, by using sub second transmission periods, there is some limit testing being done. This experiment shows that our setup is able to carry out sub second transmissions to a certain extent. Although in our end goal, we would want to have hard real-time conditions for the testbed, where a single deadline missed means system failure, this experiment shows potential and possibility in reaching that end goal.

### 5.5.12. Main Limitations of The Setup

From experiments, 5.5.6, 5.5.8 and 5.5.10, it can be seen that not having true parallel transmissions in our current setup brings about many limitations in assessing the performance. However, moving forward, we still continued to do the experiments because it still wields proper results accurate to our current system. Having true parallel transmissions would be valuable because the results it shows can be validated with our predictions based on the theoretical side of things. As we increase the size of the testbed, it would only be logical to overcome this limitation to avoid many more limitations not just in assessment but performing basic timely transmissions.

Additionally, from experiments 5.5.8 and 5.5.11, it can be observed that whenever the varying of channels were involved, the test results we get seem to go against expectations. Although the exact reason is still unclear, we can attribute it to some non-ideal channel conditions that makes the wireless medium uncertain, which are generally not in the user's control, which led to packets being missed at higher rates in the receiving gateway, resulting in anomalies in certain test results.

# 6. Future Works

## 6.1. Experimental Limitations

As encountered many times throughout the experiment, due to the program being multithreaded and the nature of SPI instructions, there is a need for locks to be used to prevent race conditions from happening. However, this also means that the "parallel" transmissions of the radios are restricted by the locks, as only one radio can have access to the lock at any given time. In order to have true parallel transmissions, perhaps the use of a field-programmable gate array (FPGA) can be explored.

Also encountered multiple times in the experiments, varying channels in a test case seem to always yield anomalies in the test results. Hence, it is important to carry out further experiments just to verify whether each frequency in the channel list is working as intended.

## 6.2. Synchronizing Multiple Blocks

This paper mainly focuses on getting one essential building block of the entire real-time LoRa testbed to work and then evaluating the performance of it. The next step to take would be to integrate multiple blocks by synchronizing the RPIs of each block via an NTP and evaluate the performance of it. In this component, there might be difficulties such as network latency which might create difficulties in creating precise synchronization between the RPIs.

## 6.3. Scalability Testing

In order to increase the size of the testbed, each RPI will have to be able to support more LoRa radios. As it stands, power will not be a concern if the radios do not transmit in parallel. This means that the limiting factor for the maximum number of radios that can be connected is the number of available GPIO pins. Since there are 28 total GPIO pins on the RPI, up to seven radios can be connected to the RPI. Any further expansions would require the use of other equipment to increase the number of accessible GPIO pins on the RPI. The performance of the setup would need to be evaluated again with a larger number of radios.

# 7. Conclusion

This paper aimed to boost the efforts put into the research on real-time LoRa communications by attempting to create a large-scale real-time LoRa testbed which has not yet been created. It also touched on the basic concepts of SPI, LoRa modulation, the difficulties and struggles of creating a large-scale real-time LoRa testbed. In order to overcome these difficulties, the paper explored the use of RPIs to communicate with LoRa radios via the SPI protocol. With the help of RadioLib, we managed to get a single RPI to communicate with three LoRa radios and transmit based on a scheduler.

From the test cases, we have observed the effects of varying the frequencies and the SFs used. The optimal values for these parameters would depend heavily on the requirements for the application. For example, in an environment with high levels of noise, a setup with high SNR would be preferred, but that would also mean a higher ToA. Although limited, we have also made theoretical comparisons to a different medium access control (MAC) protocol, CSMA, and showed that TDMA would perform better in terms of energy efficiency. Comparisons have also been made between aperiodic and periodic transmissions, which showed that periodic transmission periods have a better PLR. Important features of a real-time LoRa testbed, such as having milliseconds accuracy transmissions have also been worked towards in terms of delay jitter. Additionally, a method to derive the optimal transmission period, given a specific SF was also discussed. Lastly, the general performance of the testbed has also been evaluated in terms of the PRR across the different test cases. The limitations of our current setup which came in the form of the lack of true parallel transmissions and anomalous behaviour from varying channels were then discussed, along with the complications it brought in certain experiments.

These findings are important because they can provide practical insights such as how different parameters and MAC protocols can affect the performance of the testbed in terms of SNR, RSSI, PLR and PRR. More importantly, these results tell us that the current setup that we have is able to make transmissions with relatively high-performance metrics. As a small component of the large-scale real-time LoRa testbed that we desire, these findings play an important role by merely showing the feasibility of creating that end product. As a recap, we first brought up the question of "Can we build a scalable and easy-to-deploy real-time LoRa testbed?" Previously, there were doubts about that question because it hasn't been done before. However, after conducting this research, we know for sure that the answer to that question is now "not impossible".

With these efforts, we hope that our work has built a solid foundation that serves as the building blocks of the desired end product, cutting the time required before a working large-scale LoRa testbed can be delivered confidently to the real-time community.

[1] Molisch, A. F. (2012). *Wireless communications* (Vol. 34). John Wiley & Sons.

[2] The Things Network. (2024, March 19). What are LoRa and LoRaWAN? https://www.thethingsnetwork.org/docs/lorawan/what-is-lorawan/

[3] Bor, M., & Roedig, U. (2017). LoRa transmission parameter selection. *In 2017 13th International Conference on Distributed Computing in Sensor Systems (DCOSS)* (pp. 27–34). IEEE. https://doi.org/10.1109/DCOSS.2017.10

[4] Devalal, S., & Karthikeyan, A. (2018). LoRa technology - An overview. *In 2018 Second International Conference on Electronics, Communication and Aerospace Technology (ICECA)* (pp. 284–290). IEEE. https://doi.org/10.1109/ICECA.2018.8474715

[5] Maleki, A., Nguyen, H. H., Bedeer, E., & Barton, R. (2024). A tutorial on chirp spread spectrum modulation for LoRaWAN: Basics and key advances. *IEEE Open Journal of the Communications Society, 5*, 4578–4612. https://doi.org/10.1109/OJCOMS.2024.3433502

[6] Dhaker, P. (2018). Introduction to SPI interface. *Analog Dialogue, 52*(3), 49–53.

[7] Leens, F. (2009). An introduction to I2C and SPI protocols. *IEEE Instrumentation & Measurement Magazine, 12*(1), 8–13. https://doi.org/10.1109/MIM.2009.4762946

[8] Dhanalaxmi, B., & Naidu, G. A. (2017). A survey on design and analysis of robust IoT architecture. *In 2017 International Conference on Innovative Mechanisms for Industry Applications (ICIMIA)* (pp. 375–378). IEEE. https://doi.org/10.1109/ICIMIA.2017.7975639

[9] Gridling, G., & Weiss, B. (2007). *Introduction to microcontrollers*. Vienna University of Technology Institute of Computer Engineering Embedded Computing Systems Group, 25.

[10] Jgromes. (n.d.). JGROMES/RadioLib: Universal wireless communication library for embedded devices. GitHub. https://github.com/jgromes/RadioLib

[11] Ebyte. (n.d.). *E22-900M30S SX1262 LoRa module*. Chengdu Ebyte Electronic Technology Co., Ltd. https://www.cdebyte.com/products/E22-900M30S/1#Parameters

[12] Raspberry Pi. (n.d.). *Raspberry Pi documentation*. https://www.raspberrypi.com/documentation/computers/raspberry-pi.html

[13] Yoo, S., & Jerraya, A. A. (2003). Introduction to hardware abstraction layers for SoC. In A. A. Jerraya, S. Yoo, D. Verkest, & N. Wehn (Eds.), *Embedded software for SoC* (pp. 453–467). Springer. https://doi.org/10.1007/0-306-48709-8_14

[14] Lgpio C API (local). (n.d.). lg library. https://abyz.me.uk/lg/lgpio.html

[15] Vittoz, E. A., Degrauwe, M. G. R., & Bitz, S. (1988). High-performance crystal oscillator circuits: Theory and application. *IEEE Journal of Solid-State Circuits, 23*(3), 774–783. https://doi.org/10.1109/4.318

[16] Mills, D. L. (1991). Internet time synchronization: The network time protocol. *IEEE Transactions on Communications, 39*(10), 1482–1493. https://doi.org/10.1109/26.103043

[17] Chapre, Y., Mohapatra, P., Jha, S., & Seneviratne, A. (2013). Received signal strength indicator and its analysis in a typical WLAN system (short paper). In *38th Annual IEEE Conference on Local Computer Networks* (pp. 304–307). IEEE. https://doi.org/10.1109/LCN.2013.6761255

[18] Semtech. (2013, July 1). *SX1272/3/6/7/8: LoRa modem designer's guide AN1200.13*.

[19] Bentem, A. (2020, July 21). Airtime calculator for LoRaWAN. https://avbentem.github.io/airtime-calculator/

[20] "Real-Time Communication over LoRa Network", *Proc. IEEE Int. Conf. Internet-of-Things Design and Implementation (IoTDI)*, pp. 14–27, 2022

[21] Gamage, A., Liando, J. C., Gu, C., Tan, R., & Li, M. (2020). LMAC: Efficient carrier-sense multiple access for LoRa. *Proceedings of the 26th Annual International Conference on Mobile Computing and Networking*, 1–13.